\documentclass{iau} 

\usepackage{graphicx}
\usepackage{xcolor}
\usepackage[hyphens]{url}
\usepackage{hyperref}
\hypersetup{
     colorlinks  = true,
     linkcolor   = blue, 
     anchorcolor = BrickRed,
     citecolor   = blue,
     filecolor   = blue,
     menucolor   = blue,
     runcolor    = cyan,
     urlcolor    = blue
}
\usepackage[authoryear]{natbib}

\pubyear{2019}
\volume{355} 
\jname{The Realm of the Low-Surface-Brightness Universe}
\editors{D. Valls-Gabaud, I. Trujillo \& S. Okamoto, eds.}

\def\etal{\textit{et al.}}

\title[Up-bending breaks in galaxy disks] 
{On the origins of up-bending breaks in galaxy disks}

\author[A.~E. Watkins et al.]   
{Aaron E. Watkins$^1$,
   Jarkko Laine$^2$,
  S\'{e}bastien Comer\'{o}n$^1$,
  Joachim Janz$^{1,3}$
  \and Heikki Salo$^1$}

\affiliation{$^1$Astronomy Research Unit, University of Oulu, \\
  FIN-90014, Oulu, Finland  \\[\affilskip] 
  $^2$Hamburg Sternwarte, Universit\"{a}t Hamburg, \\ 21029,
  Hamburg, Germany  \\[\affilskip] 
  $^3$Finnish Centre of Astronomy with ESO (FINCA), University of
  Turku, \\ V\"{a}is\"{a}l\"{a}ntie 20, 21500 Piikki\"{o}, Finland  } 

\begin{document}

\maketitle

\begin{abstract}
Using \emph{SPITZER} 3.6$\mu$m imaging, we investigate the physical
and data-driven origins of up-bending (Type III) disk breaks.  We
apply a robust new break-finding algorithm to 175 low-inclination disk
galaxies previously identified as containing Type III breaks, classify
each galaxy by its outermost re-classified (via our new algorithm)
break type, and compare the local environments of each resulting
subgroup.  Using three different measures of the local density of
galaxies, we find that galaxies with extended outer spheroids (Type
IIIs) occupy the highest density environments in our sample, while
those with extended down-bending (Type II) disks and symmetric
outskirts occupy the lowest density environments.  Among outermost
breaks, the most common origin of Type III breaks in our sample is
methodological; the use of elliptical apertures to measure the radial
profiles of asymmetric galaxies usually results in features akin to
Type III breaks.
\end{abstract}

\firstsection 

\section{Context}

It has been known for some time that the surface brightness profiles
of disk galaxies are only approximately exponential, insofar as most
show at least one prominent and lasting change in slope at some radius
\citep[e.g.,][]{freeman70, pohlen06}.  As surface brightness profiles
trace the radial distribution of stellar mass (modulo the mass-to-light
ratio), these changes---or breaks---in slope seem closely tied to the
evolution of their hosts.  Precisely what aspects of disk galaxy
evolution they illustrate, however, remains somewhat ambiguous.

Up-bending \citep[or Type III;][]{pohlen06} breaks specifically may
have numerous origins, from accretion of dwarf companions
\citep[e.g.,][]{younger07} to star formation \citep[e.g.,][]{laine16}
to scattering by bars \citep[e.g.,][]{herpich17}.  If so, simply
identifying whether or not a galaxy hosts a Type III break, without
additional distinctions, may not be physically illuminating.  While
past studies have identified several sub-classifications for Type III
breaks \citep[e.g.,][]{pohlen06, erwin08}, which might be useful in
making such distinctions, most avoid considering these
sub-classifications as separate populations.  On top of this, disk
break studies suffer from a repeatability problem; only $2/3$ of
galaxies show matching classifications across studies \citep{laine16},
leading to further ambiguity.

We have therefore devised a robust and nearly hands-off break-finding
algorithm, which we applied to a sample of Type III break-host
galaxies \citep[as previously identified by][]{laine14, laine16}.  We
find that meaningful physical trends can be uncovered through careful
classification of Type III breaks even with a small sample of
galaxies.

\section{Method}

We designed our break-finding algorithm to minimize the need for
decision making.  One potential source of confusion in past studies is
the need to differentiate between breaks, being lasting changes in
slope, and smaller scale bumps or wiggles in the profile \citep[see,
  e.g., the Appendix of][]{pohlen06}; therefore, we chose to make no
such distinction.  Instead, we measured the slope of the surface
brightness profile as a function of radius in the manner detailed by
\citet{pohlen06}, and applied a method called change-point analysis
\citep{taylor00} to determine each point where the slope changes.

We then classified each break in detail.  We found that Type III
breaks arise in three circumstances: where the isophotes become
consecutively rounder with increasing radius \citep[Type IIIs, denoted
  Type III-h by][]{pohlen06}; where the photometric apertures
measuring the profile first encounter a ring, set of spiral arms, or
other local light enhancement \citep[Type IIId, denoted Type III-d
  by][]{pohlen06}; or where these photometric apertures encounter a
region in which the disk is asymmetric, whether through a global
asymmetry, tidal debris, or otherwise (Type IIIa).  The latter should
likely not be considered breaks, as the shape of the radial profile of
an asymmetric galaxy is simply not reliable, yet aside from Type IIId
these are the most numerous break type in our sample.  Many galaxies
in our sample also contained Type II (down-bending) breaks, hence
depending on their origins we classified them as either Type IId
(arising due to association with disklike features such as spiral
arms) or IIa \citep[arising due to asymmetry, akin to Type II-AB
  from][]{pohlen06}.  Finally, in some cases we found no significant
change points across the disk (Type I), and in a handful of cases we
found that the disk profiles were better described using a higher
S\'{e}rsic index model (Type 0).

To investigate the impact of environment on the formation of these
various break types, we classified our sample galaxies by their
outermost breaks, these being the most sensitive to environmental
influences.  We then compared populations of galaxies using three
different environmental parameters: $Q$, the Dahari tidal parameter
\citep{dahari84}; $\Sigma_{3}^{A}$, the projected surface density of
galaxies within the third-nearest neighbor distance
\citep[e.g.,][]{cappellari11}; and $\Sigma_{\rm KT17}^{A}$, defined as
$\Sigma_{3}^{A}$ but using the group radii given by \citet{kourkchi17}
in lieu of the third-nearest neighbor distance.  We did these
comparisons pair-wise, using the BEST method \citep[Bayesian
  estimation supersedes the \emph{t} test;][]{kruschke13} to compare
the mean values of each environmental parameter for each group of
galaxies.  For a full description of our methodology, see
\citet{watkins19}.

\section{Results}

\begin{figure}[b]
\begin{center}
 \includegraphics[width=0.7\textwidth]{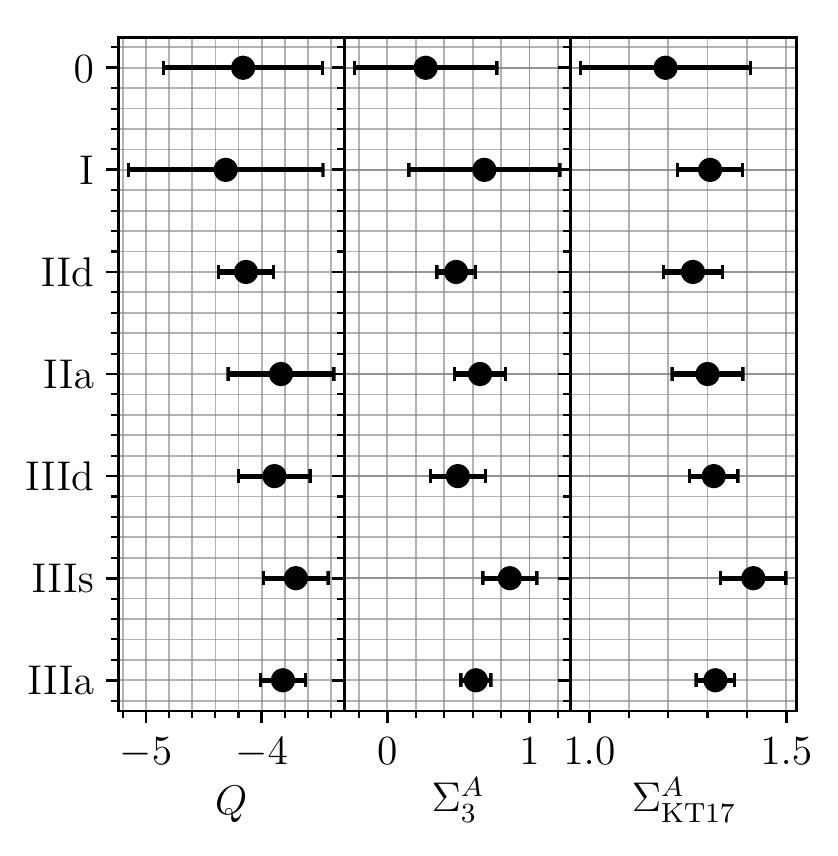} 
 \caption{Comparisons in local density of neighboring galaxies between
   sub-samples of our Type III galaxies.  Our sample is separated by
   outermost break classification (excluding Types I and 0, which were
   not found to have significant breaks).  See text for description of local
   environmental parameters and break sub-classifications.
  \label{fig:scatter}}
\end{center}
\end{figure}

Our results are summarized in Fig.\,\ref{fig:scatter}.  Despite the
fairly small number of galaxies in our sample, made even smaller
through separation into seven sub-categories, we found robust trends
between outermost break type and local environmental density.
Specifically, galaxies with outer spheroids (Type IIIs) occupy the
highest density environments, while those host to Type II breaks
aligned with symmetric disk features (Type IId) occupy the lowest
density environments, despite most such galaxies hosting Type III
breaks at smaller radii.  Also, asymmetric galaxies, regardless of
break direction (those ending with Type IIa and IIIa breaks) occupy
similarly dense environments, slightly lower in density than those of
Type IIIs.

These trends paint a consistent physical portrait.  Galaxies in
high-density environments (e.g., compact groups or clusters) may more
frequently suffer harrassment \citep{moore96}, leading to thermal
excitation of their stellar populations, visible as an outer spheroid
component.  Galaxies in slightly lower density environments
(e.g., loose groups) will be tidally disturbed, but less frequently and
by more prolonged tidal interactions, leading to more prominent tidal
tails and other signs of asymmetry \citep[e.g.,][]{barnes92}.
Finally, galaxies in the least dense environments will suffer little
tidal interaction, leaving their structure to be dominated by secular
processes such as the formation of symmetric spiral arms.

In total, then, our method is able to tease out intriguing physical
trends from disk breaks for even a small population of galaxies.  It
will be of interest to apply such a method to a much larger sample,
both to verify these trends and to possibly uncover subtler physical
processes over a wider variety of environments.  Because the
break-finding routine requires so few assumptions, it can be easily
automated, making it adaptable to large-scale surveys.  Detailed
classifications, while currently done by-eye, can also be automated to
some extent given that properties such as asymmetry are quantifiable.
This method therefore is quite promising for future studies of the
radial mass distribution of disk galaxies.


\end{document}